\pdfoutput=1
\documentclass{JINST}

\usepackage{subcaption}
\usepackage{amstext}

\title{\boldmath Effect of glass thickness variations on the performance of RPC detectors}

\author{J. Sadiq$^{a}$, K. Raveendrababu$^{a,b}$, and P. K. Behera$^{a}$ \\
\llap{$^{a}$}Indian Institute of Technology, Madras, Chennai, India\\
\llap{$^{b}$}Physical Science Division, Homi Bhabha National Institute, Anushaktinagar, Mumbai, India \\
E-mail: \email{sadiqvengad@gmail.com}}

\abstract
{
The India-based Neutrino Observatory (INO) is planning to build a magnetized iron calorimeter detector (ICAL) 
in which Resistive Plate Chambers (RPCs) will be the active detector elements.
A study of the performance of RPCs, made using electrodes of various thicknesses, is pivotal in optimizing the design parameters of the ICAL RPCs.
We fabricated RPCs with glasses of various thicknesses and studied their performance in the same environmental conditions.
A study of detector efficiency, noise rate, time resolution and charge distribution is presented in this paper.
}


\begin{document}
\maketitle
\flushbottom

\section{Introduction}
\label{sec:intro}

Resistive Plate Chambers (RPCs) have widespread acceptance in the detection of charged particles 
because of their excellent time resolution \cite{santonico_rpcdev}. 
Good performance of RPCs is reported from many particle physics experiments \cite{rpc_CMS_perfomance, OPERA_rpc, ATLAS_rpc}. 
The India-based Neutrino Observatory (INO) is a proposed underground laboratory 
which will house a large magnetised iron calorimeter (ICAL) detector 
to accurately measure the parameters related to the atmospheric neutrino oscillations.
RPCs will be used as active detector elements in the INO-ICAL detector \cite{icalwhitepaper}.
The ICAL will consist of 151 layers of 5.6 cm thick iron plates interleaved by a 4 cm air gap containing the RPCs.
About 28,800 glass RPCs of 2 m $\times$ 2 m size will be used in the ICAL detector consisting of three modules,
each of size 16 m $\times$ 16 m $\times$ 14.5 m.

RPCs are gaseous detectors in which a constant and uniform electric field, produced by two parallel plate electrodes of high bulk resistivity,
is applied to the gas gap. The propagation of the charges, produced by the ionization of an incoming charged particle, 
induces a current signal on external pickup strips \cite{RPC_simulation_1}.
The electric field inside the gas gap and hence the charge induced on the strips depend on the detector parameters 
like the thickness and permittivity of both the glass and the gap \cite{rpcfield}.
We have studied the effect of thickness variations of the glass electrodes on the performance of RPCs.
Efficiency, noise rate, charge distribution and time resolution of the RPCs were studied.

\section{Experimental setup}

Glass plates of thickness 3.24 mm, 3.97 mm and 4.91 mm (which are referred as 3 mm, 4 mm and 5 mm in this paper), manufactured by Saint-Gobain, 
were cut in 30 cm $\times$ 30 cm dimensions.
A thin conductive tape (T-9149) of uniform surface resistance was pasted on the outer surfaces of the glass plates.
These electrodes were used to fabricate 2 mm gap RPCs. 
The RPCs were operated in avalanche mode with a gas mixture of R134a ($C_2H_2F_4$, 95$\%$), Isobutane ($C_4H_{10}$, 4.5$\%$) 
and Sulphur hexafluoride ($SF_6$, 0.5$\%$).
The RPCs were sandwiched between two pick-up panels consisting of 10 copper strips each of width 28 mm, separated by a gap of 2 mm.
The central strip was read out to study the performance of RPCs.

\begin{figure}[htbp]
      \hspace{-5mm}
      \centering
      \includegraphics[width=0.95\linewidth,trim = 15mm 20mm 0mm 85mm, clip]{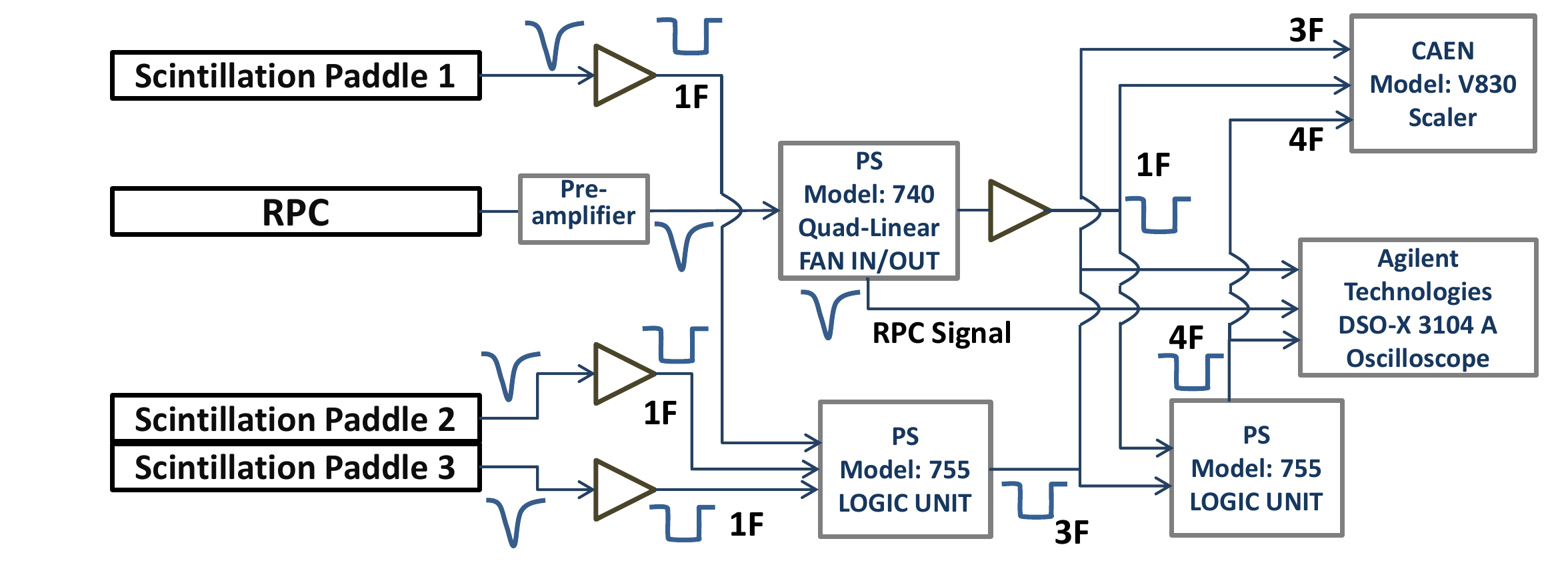}
      \vspace{-2mm}
      \caption{Schematic of experimental setup.}
      \vspace{-5mm}
      \label{fig:exp_setup}
    \end{figure}

The RPCs were tested using cosmic ray muons. A cosmic ray telescope was made using three plastic scintillator paddles P$_1$, P$_2$ and P$_3$,
arranged vertically one above the other.
The RPC was placed in such a way that the device under study was sandwiched between the scintillator paddles.
Schematic of experimental setup is shown in figure \ref{fig:exp_setup}.
The analog signals from the RPC were amplified with a pre-amplifier, since the RPC signals are small in amplitude in avalanche mode of operation.
The signals from the scintillator paddles and the amplified RPC signals were fed to the Data Acquisition System (DAQ).
Scintillator paddle signals were converted into logic signals by a discriminator.
These logic signals were ANDed in a logic unit to form the trigger pulse.
The amplified RPC analog signal was input to a linear FAN IN/FAN OUT(FIFO) to get two buffered output signals.
One output signal was fed to an oscilloscope to measure the time and charge of the signal 
and the other was converted to a logic signal by feeding it to a discriminator with a threshold voltage of -20 mV.

\section{Results and discussions}

\subsection{V-I Characteristics}

\begin{figure}[htbp]
    \vspace{-3mm}
     \begin{minipage}{0.45\textwidth}
        \centering
        \includegraphics[width=0.537\textwidth,trim = 25mm 22mm 25mm 20mm, clip]{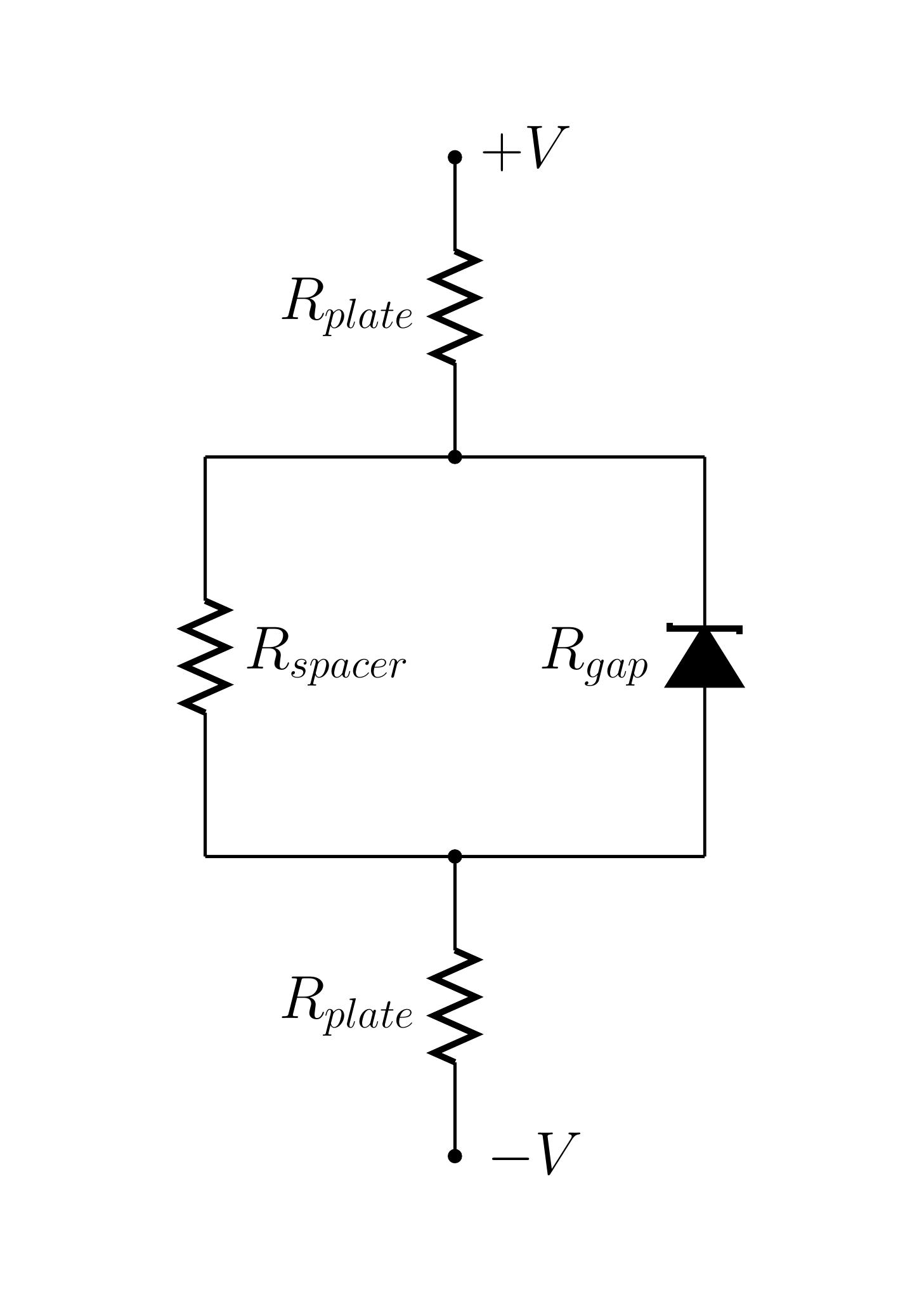}
        \caption{Equivalent circuit of an RPC.}
        \label{eq_circuit}
    \end{minipage}%
    \raisebox{0.004\textwidth}{%
    \begin{minipage}{0.55\textwidth}
        \hspace{-11mm}
        \includegraphics[width=1.14\textwidth,trim = 0mm 2mm 0mm 8mm, clip]{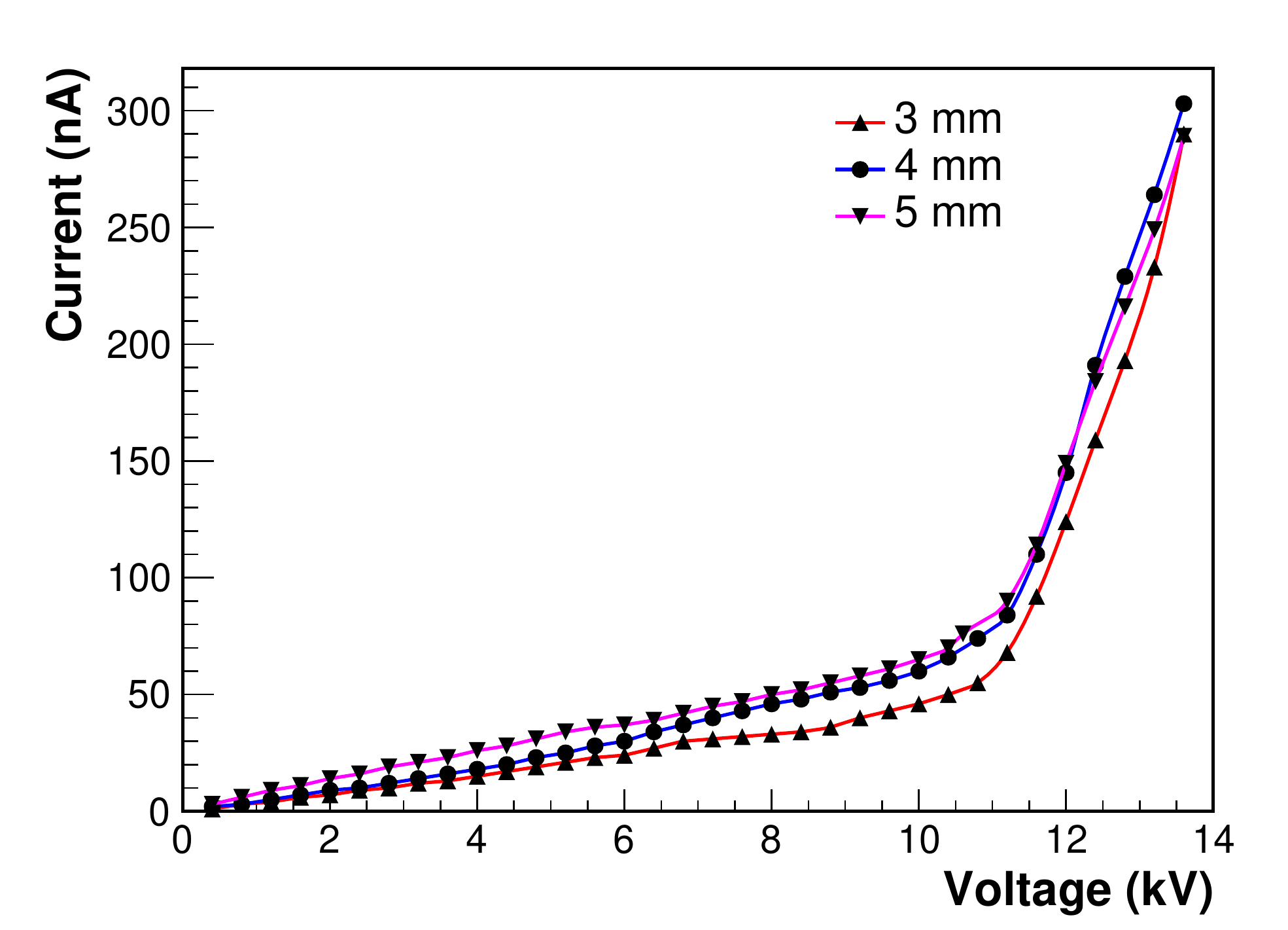}
        \caption{V-I Characteristics of the RPCs.}
        \label{VI}
    \end{minipage}
    }
    \vspace{-3mm}
  \end{figure}

To study the V-I Characteristics, we applied a varying voltage across the RPC and measured the current for each voltage.
The current depends on the applied voltage and shows a threshold behaviour.
Figure \ref{eq_circuit} represents an equivalent electric circuit for an RPC, suitable for our purposes \cite{Salim_SF6}.
If the applied voltage is sufficiently high, avalanches will be formed by multiplication of charges in the gas gap.
Hence, the gap resistance will be reduced and the current is determined by the glass resistance alone.
At lower voltages, the primary electron-ion pairs created will not produce avalanches.
Therefore, the resistance of the gap will be infinite and the current will flow through the spacer resistance.
The V-I Characteristics of RPCs for various glass thicknesses are shown in figure \ref{VI}.
There is a small difference in the current of the three RPCs below the threshold voltage.

\subsection{Efficiency}
\begin{figure}[htbp]
    \vspace{-3mm}
    \hspace{-2mm}
    \centering
    \begin{minipage}{0.49\textwidth}
        \includegraphics[width=1.05\linewidth]{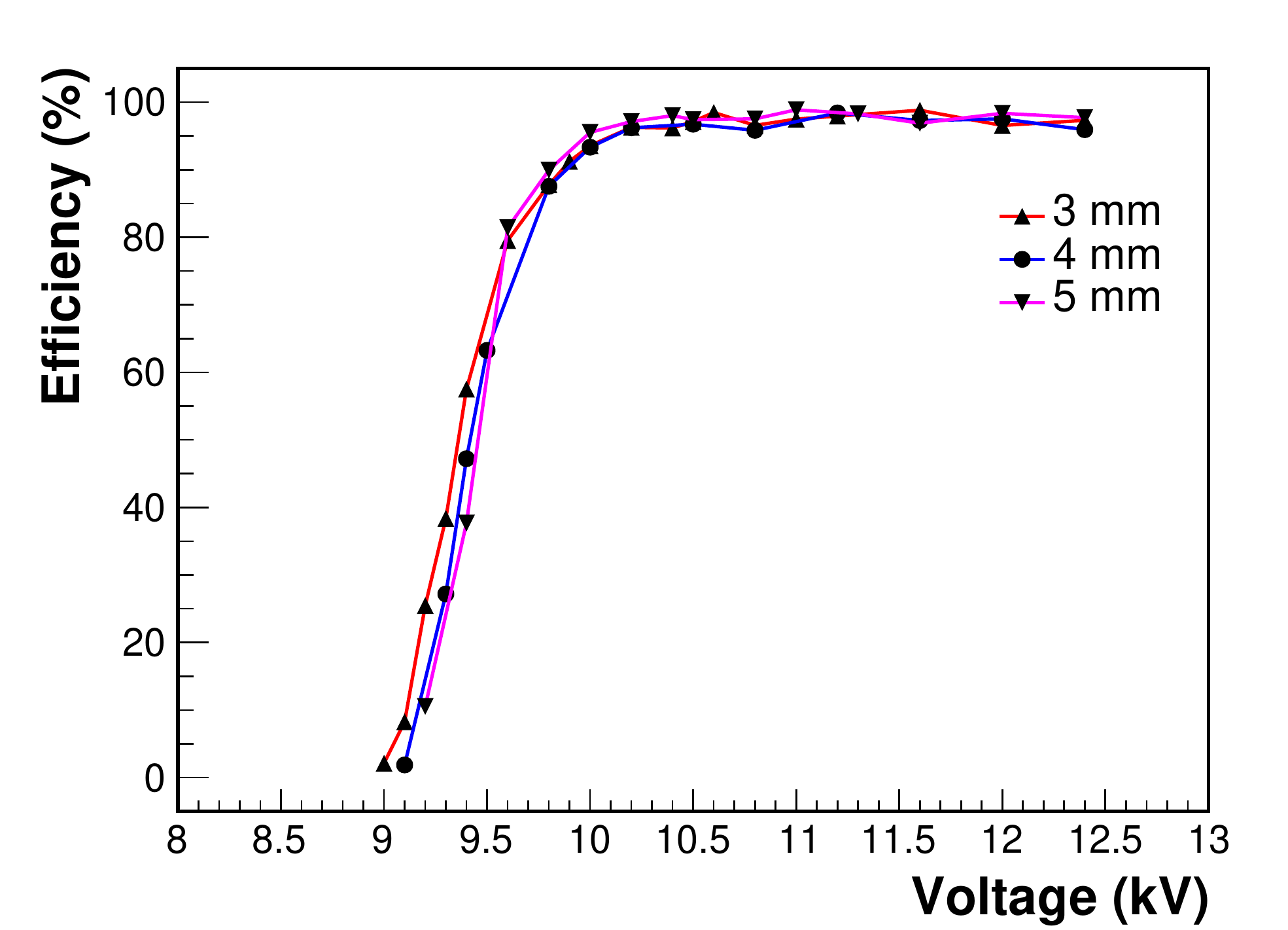}
        \caption{Efficiency of the RPCs.}
        \label{fig:efficiency}
    \end{minipage}
    \begin{minipage}{0.49\textwidth}
        \includegraphics[width=1.05\linewidth]{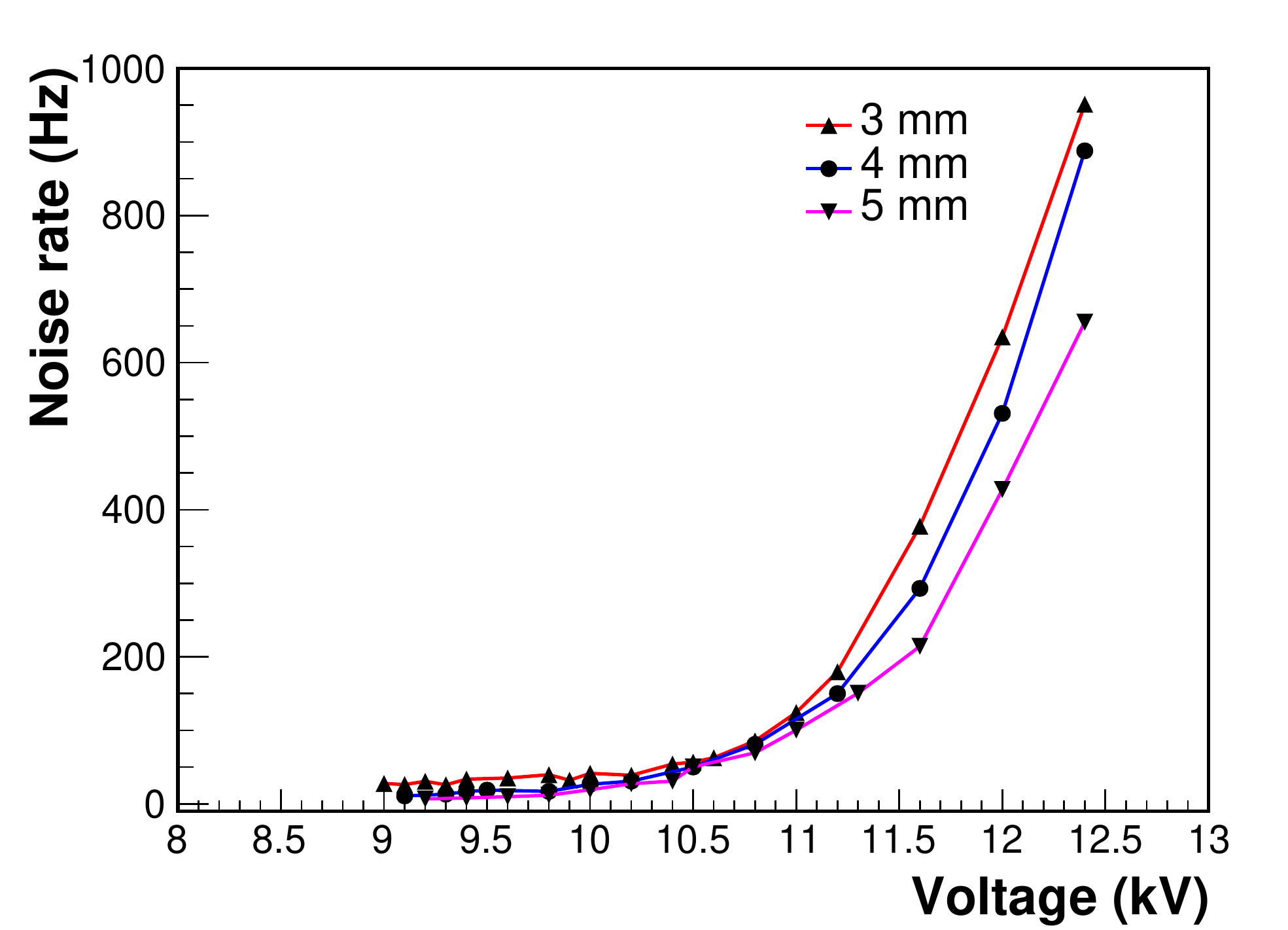}
        \caption{Noise rate of the RPCs.}
        \label{fig:noise}
    \end{minipage}
   \end{figure}

The efficiency of RPCs of various glass thicknesses was studied in the same conditions.
The efficiency was calculated using the equation:
\begin{equation}
\text{Efficiency} = \frac{4F \text{count}}{3F \text{count}} \times 100 \%,
\end{equation}
where $3F$ is the trigger signal, which is the coincidence of the signals from the three scintillator paddles, 
and $4F$ is the coincidence of the trigger signal and the RPC strip signal.
Figure \ref{fig:efficiency} shows the efficiency as a function of applied voltage.
Efficiency increases with voltage and gets saturated at higher voltages. 
At low applied voltages (9 - 9.5 kV), the efficiency of 3 mm RPC is slightly higher with respect to the other two RPCs.
This is presumably due to the fact that the charge induced on the strip with RPCs of lower thickness electrodes is comparatively more
as discussed in section \ref{QandT}. Therefore, even the muons, which create less number of charge pairs in the gap, can give a signal above threshold.
The asymptotic efficiency ($\varepsilon_{max}$) was found to be more than 95$\%$ for all RPCs.
The knee voltage is the value for which 95$\%$ of the asymptotic efficiency is reached \cite{CMSRPC}.
The knee voltage was found to be same (10.0 kV) in the three cases considered.

\subsection{Noise rate of the strip}

Noise rate is the measure of the random signals in the strip, 
which are produced by stray radio activity and micro-discharges of the chamber.
Figure \ref{fig:noise} shows the noise rate as a function of applied voltage.
An increase in the applied electric field leads to an increase in the noise rate.
Once the knee voltage is reached, the noise rate starts to increase sharply.
Particles with very low energy will lose their kinetic energy before reaching the gas gap.
An increased number of low energy particles will be blocked when thicker glass electrode is used.
This could be the reason for getting comparatively lower noise rate for RPCs with thicker electrodes.
Moreover, small signals inside the gap would induce signals under threshold on read-out electrodes when the electrode thickness is higher, 
due to the same reason outlined for the efficiency.

\subsection{Charge distribution}
\label{QandT}

    \begin{figure}[htbp]
        \hspace{-6mm}
        \centering
        \begin{subfigure}[b]{0.49\textwidth}
                \includegraphics[width=1.05\textwidth,trim = 0mm 0mm 0mm 0mm, clip]{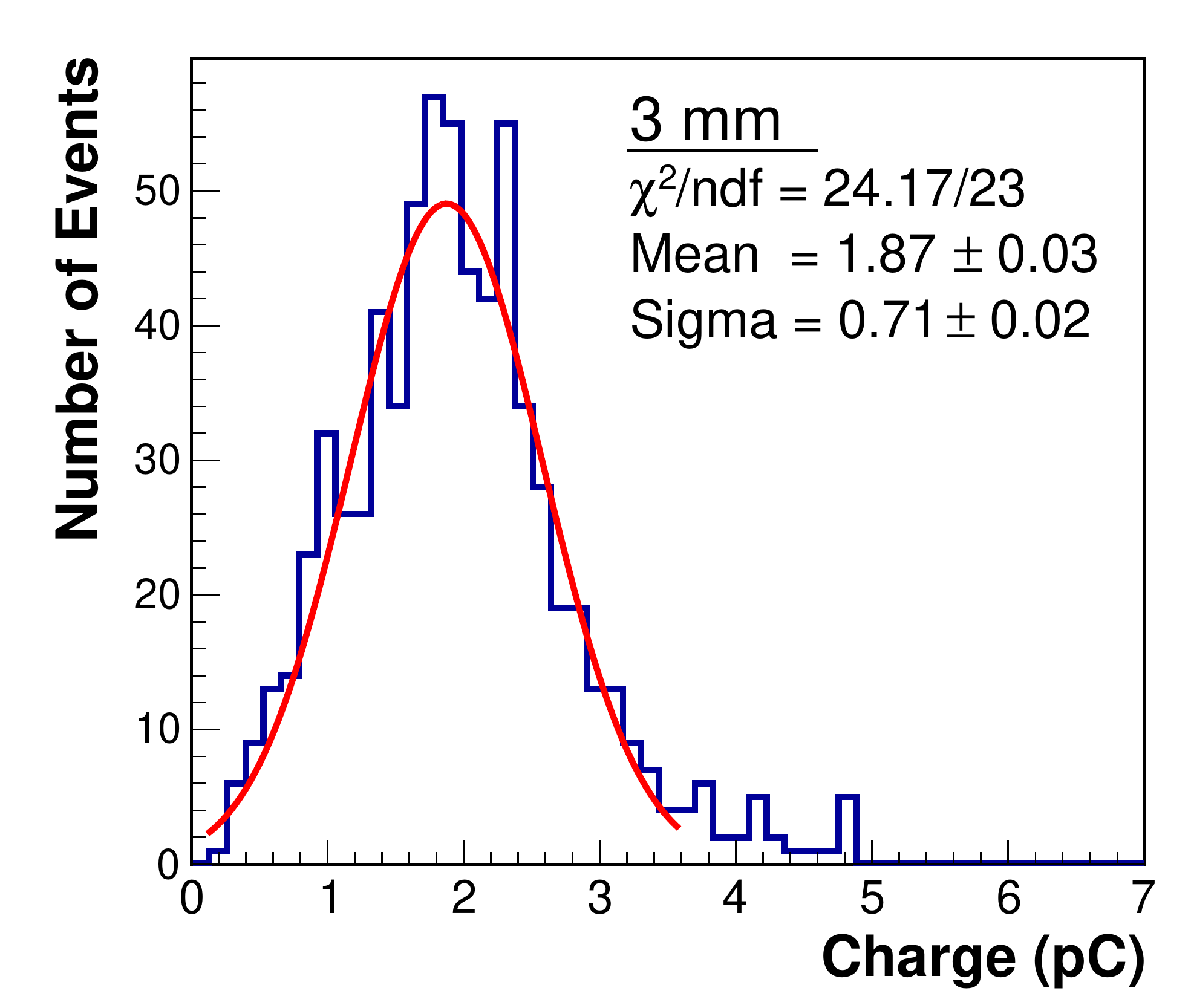}
        \end{subfigure}
        \begin{subfigure}[b]{0.49\textwidth}
                \includegraphics[width=1.05\textwidth,trim = 0mm 0mm 0mm 0mm, clip]{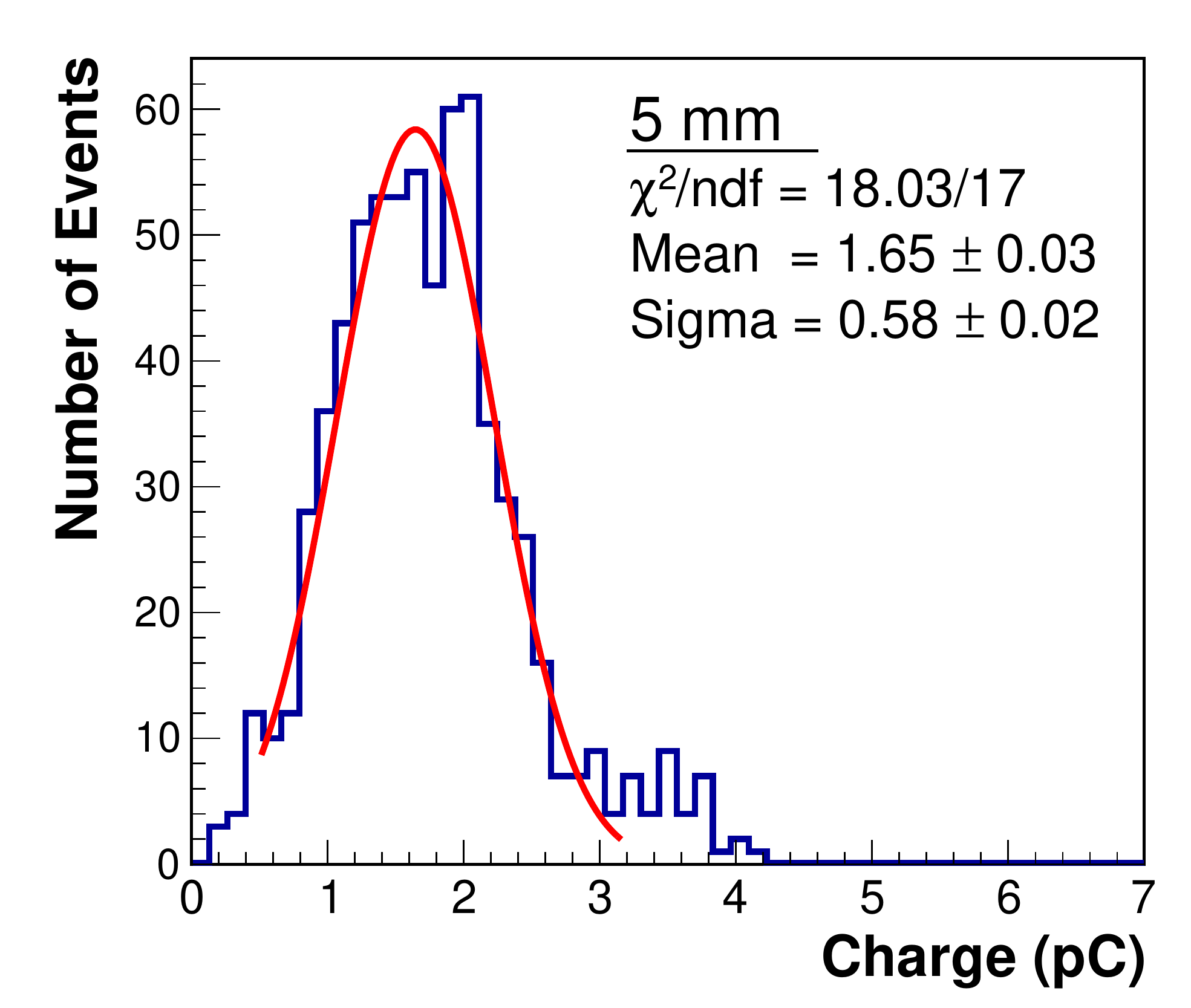}
        \end{subfigure}
        \caption{Charge distribution of the RPCs with glass thicknesses 3 mm and 5 mm at 10.5 kV.}
        \label{fig:charge}
    \end{figure}

The charge induced on the strip was measured using an oscilloscope (DSO-X 3104 A) which stored the digitized analog pulses. 
The charge distributions of 700 signals collected with RPCs of glass thickness 3 mm and 5 mm were fitted with Gaussians
as shown in figure \ref{fig:charge}. The mean charge was found to be higher for the RPC of lower electrode thickness.
This can be explained by the fact that if $Q$ is the total charge delivered in the gas, the charge induced on the strip is given by,
  \begin{equation}\label{charge}
   Q_{ind} = Q/\Bigg(1+\frac{2d}{\varepsilon_rg}\Bigg)
  \end{equation}
where $g$ is the gap size and $d$ is the thickness of electrode having relative permittivity $\varepsilon_r$ \cite{RPC_simulation_1}.
The attenuation factor ($A = 1+2d/\varepsilon_rg$) has more importance for RPCs made with electrodes of low relative permittivity.
The relative permittivity of Saint-Gobain glass is 11 \cite{RPC_raveendra}.
So the value of $A$ for RPCs of glass thicknesses 3.24 mm and 4.91 mm is 1.29 and 1.45, respectively. 
Thus the mean charge observed for 3 mm and 5 mm RPCs are consistent with equation \ref{charge}.
    
\subsection{Time resolution}
    
\begin{figure}[htbp]
        \hspace{-2mm}
        \begin{subfigure}[b]{0.49\textwidth}
                \vspace{-2mm}
                \includegraphics[width=1.05\textwidth,trim = 0mm 0mm 0mm 0mm, clip]{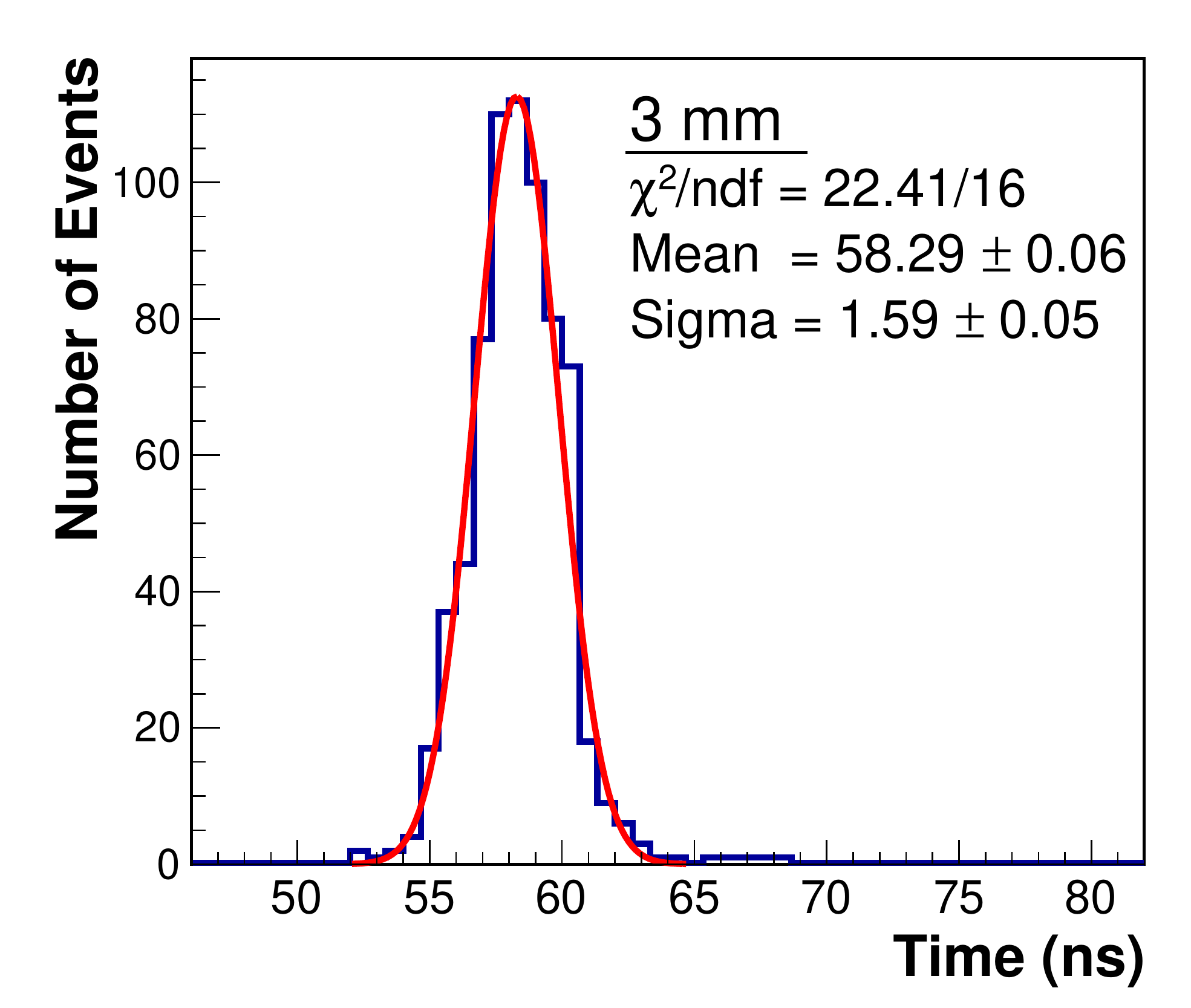}
                \vspace{-8mm}
        \end{subfigure}
        \begin{subfigure}[b]{0.49\textwidth}
                \vspace{-2mm}
                \includegraphics[width=1.05\textwidth,trim = 0mm 0mm 0mm 0mm, clip]{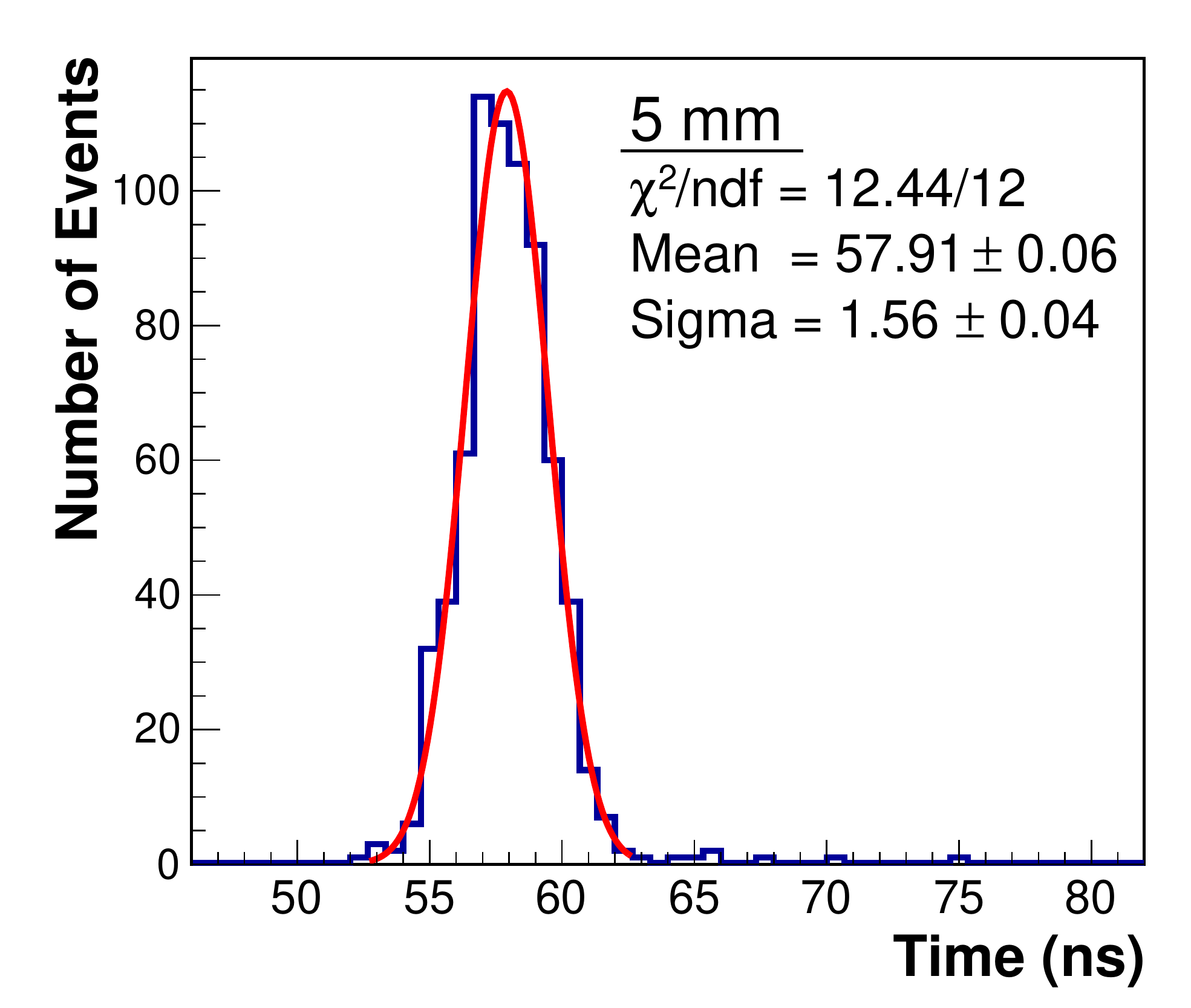}
                \vspace{-8mm}
        \end{subfigure}
        \caption{Time distribution of the RPCs with glass thicknesses 3 mm and 5 mm at 10.5 kV.}
        \label{fig:time}
        \vspace{-3mm}
    \end{figure}

The time distributions for the RPCs made using 3 mm and 5 mm electrodes were measured for 700 signals and are shown in figure \ref{fig:time}.
A time resolution of about 1.6 ns was obtained for both the RPCs when they were being operated at 10.5 kV.

\section{Conclusions}

Performance of RPCs made using glass electrodes of three different thicknesses was studied.
There is no significant difference in the knee voltage of the three RPCs studied.
The noise rate can be reduced to some extent by using thicker glass electrodes.
The charge induced on the strip is found to be higher for the RPC of lower electrode thickness, 
since the charge is reduced by a factor which depends on the ratio of the electrode thickness to the gap size.
The time resolution was found to be similar for both the RPCs.

\section*{Acknowledgements}

This work was done with the support of the Department of Atomic Energy (DAE), and the Department of Science and Technology (DST), Government of India.
The authors would like to thank R. R. Shinde, S. D. Kalmani at TIFR, Mumbai and V. Janarthanam at IIT Madras for their help throughout the work.

\end{document}